\pdfoutput=1
\def\year{2020}\relax
\documentclass[letterpaper]{article} 
\usepackage{aaai20}  
\usepackage{times}  
\usepackage{helvet} 
\usepackage{courier}  
\usepackage[hyphens]{url}  
\usepackage[dvipdfmx]{graphicx}
\usepackage{bmpsize}
\usepackage{cite}
\usepackage{amsmath,amssymb,amsfonts}
\usepackage{algorithmic}
\usepackage{graphicx}
\usepackage{textcomp}
\usepackage{bmpsize}
\usepackage{xcolor}
\usepackage{lipsum}
\usepackage[colorlinks=true,urlcolor=black]{hyperref}
\usepackage{natbib}

\usepackage{amsthm}
\theoremstyle{definition}
\newtheorem{definition}{Definition}

\usepackage{algorithmic}
\usepackage{algorithm}%

\usepackage{color}
\usepackage{comment}
\usepackage{url}
\usepackage{subcaption}
\usepackage{upgreek}
\title{FLAP - A Federated Learning Framework for\\ Attribute-based Access Control Policies}
\author{Amani Abu Jabal\textsuperscript{\rm 1}, Elisa Bertino\textsuperscript{\rm 1}, Jorge Lobo\textsuperscript{\rm 2}, Dinesh Verma\textsuperscript{\rm 3}, \\\Large \textbf{Seraphin Calo\textsuperscript{\rm 3}, Alessandra Russo\textsuperscript{\rm 4}}\\
\textsuperscript{\rm 1}Department of Computer Science, Purdue University, West Lafayette, IN, USA\\
\textsuperscript{\rm 2}ICREA - Universitat Pompeo Fabra, Spain\\
\textsuperscript{\rm 3}IBM TJ Watson Research Center, Yorktown Heights, NY, USA\\
\textsuperscript{\rm 4}Imperial College, London, UK\\
\textsuperscript{\rm 1}\{bertino,aabujaba\}@purdue.edu, \textsuperscript{\rm 2}jorge.lobo@upf.edu, \textsuperscript{\rm 3}\{dverma, scalo\}@us.ibm.com, \textsuperscript{\rm 4}arusso@imperial.ac.uk
}
 
\begin{document}
\sloppy

\maketitle

\begin{abstract}
Technology advances in areas such as sensors, IoT, and robotics, enable new collaborative applications (e.g., autonomous devices). A primary requirement for such collaborations is to have a secure system which enables information sharing and information flow protection. Policy-based management system is a key mechanism for secure selective sharing of protected resources. However, policies in each party of such a collaborative environment cannot be static as they have to adapt to different contexts and situations. One advantage of collaborative applications is that each party in the collaboration can take advantage of knowledge of the other parties for learning or enhancing its own policies. We refer to this learning mechanism as \textit{policy transfer}. The design of a policy transfer framework has challenges, including policy conflicts and privacy issues. Policy conflicts typically arise because of differences in the obligations of the parties, whereas privacy issues result because of data sharing constraints for sensitive data. Hence, the policy transfer framework should be able to tackle such challenges by considering minimal sharing of data and support policy adaptation to address conflict. In the paper we propose a framework that aims at addressing such challenges. We introduce a formal definition of the policy transfer problem for attribute-based policies. We then introduce the transfer methodology that consists of three sequential steps. Finally we report experimental results.
\end{abstract}
\section{Introduction}

Recent policy-based management systems are attribute-based (AB). In these systems, policy rules are expressed as conditions against domain-meaningful properties of subjects, resources, actions, and environments. This approach simplifies policy administration as policy decisions that automatically adapt between requests based on changes of attribute values. Such a capability is critical in many collaborative applications to enhance the autonomy of the collaborating  parties, such as for example in military multi-domain operation (MDO)~\cite{ArmyTradoc}. In coalition MDO, coalition parties operating in the land, air, sea, or cyber will come together to achieve collective goals by sharing multiple viewpoints about emerging situations. Since coalition MDO contains multiple parties and types of resources, approaches to simplify policy specifications and a systematic approach to autonomously adapt policies according to the context will be critical. A major challenge is the specification of the AB policies representing the key input for policy enforcement. Since in MDO, we may typically deal with local contexts and situations, the needed detailed knowledge may be lacking. Addressing this challenge requires a distributed intelligence approach for policy learning that is able to: (i) combine datasets available at coalition parties (e.g., directories, organizational charts, logs, and existing local policies); and (ii) use machine learning (ML) to infer AB policies from these combined data.

In coalitions, parties can each have their own datasets, and combining these datasets can enhance the learning outcomes. In some cases, coalition members may only share their own local policies but not the data they used to learn their policies. In practice, a combination of those cases (i.e., sharing datasets, sharing policies) may occur. A federated approach is thus required for learning policies from a broad variety of data and knowledge, including raw data, policies expressed as rules, and ML models. It is, therefore, critical to develop an AB policy learning framework able to learn from multiple data sources while at the same time assuring that each party can generate accurate policies. To the best of our knowledge, there are no existing frameworks addressing such requirement. Recently, we have proposed a learning framework~\cite{Polisma} to learn policies of interest to a single party that uses only the data of that party. Therefore, such a framework needs to be extended in a way that enables its utilization in such a federated environment.

Towards enabling the learning framework to accommodate the differences that might be encountered in a federated environment, one main issue is the conflicts that might arise in the process of interchanging the policies between different coalition parties. Those conflicts are expected as a result of the regulation differences and security specifications. Thus, to address this issue, a similarity analysis, as well as qualitative analysis, should be performed by the target party to ensure the correctness and accuracy of the learning and transfer process. Another issue is the timeline and degree of the interaction between the policies of a pair of parties to perform a better learning and interoperability process. Therefore, we propose four approaches with different levels of interaction aiming to find the best learning output.

The paper is organized as follows. We first provide background information on ABAC policies, and introduce several definitions underlying the problem of transferring ABAC policies. We then introduce our proposed methodology and approaches, followed by  experimental results. We conclude with a discussion of related work and future work.
\section{Background and Problem Description}
\label{sec:prelim}
In this section, we first introduce background notions about  the attribute-based access control (ABAC) model. We then briefly describe our approach to learn ABC policies. Finally we formally introduce the problem addressed in this paper.
\subsection{ABAC Policies}
We assume a formal attribute-based access control (ABAC) model~\cite{xu2014mining} that includes several finite sets: users $\mathcal{U}$, resources $\mathcal{R}$, operations $\mathcal{O}$, and rules $\mathcal{P}$. Each user (i.e., $u \in \mathcal{U}$) and resource (i.e., $r \in \mathcal{R}$) is represented by independent sets of attributes (referred to as user attributes $A_U$ and resource attributes $A_R$, respectively). The values of these attributes are represented by a function which assigns to each attribute a value from the value range for the attribute as shown in expressions~\ref{eq:usr_att_val} and~\ref{eq:rsc_att_val}. An ABAC policy consists of a set of rules $\mathcal{P}$ where each rule $\rho$ is defined as $\langle e_U, e_R, O, d \rangle$ where $e_U$ is a user attribute expression, $e_R$ is a resource attribute expression, $O \subseteq \mathcal{O}$ is a set of operations, and $d$ is the decision of the rule ($d$ $\in$ $\{permit, deny\}$). An attribute expression comprises a set of attribute/value pairs. For example, a user $u_i$ satisfies $e_U$ (denoted by $u_i \models e_U$) iff for every user attribute $a$ not mapped to $\perp$, $\langle a$, $d_U({u}_{i},a)\rangle \in e_U$, and a resource $r_i$ satisfies $e_R$ (i.e., it belongs to the set defined by $e_R$, denoted by $r_i \models e_R$) iff for every resource attribute $a$ not mapped to $\perp$, $\langle a$, $d_R({r}_{i},a)\rangle \in e_R$.

{
\fontsize{9.0}{7.2}\selectfont
\begin{align}
\centering
\label{eq:usr_att_val}
d_U(u_i, a_j) &= v_k \mid u_i \in \mathcal{U} \land  a_j \in \mathcal{A_U} \land  v_k \in \mathcal{V_U}(a_j)\\
\label{eq:rsc_att_val}
d_R(r_i, a_j) &= v_k \mid r_i \in \mathcal{R} \land  a_j \in \mathcal{A_R} \land  v_k \in \mathcal{V_R}(a_j)
\end{align}
}

\subsection{Policy Learning}
\begin{figure*}[!htbp]
	\centering
    \includegraphics[width=0.6\linewidth]{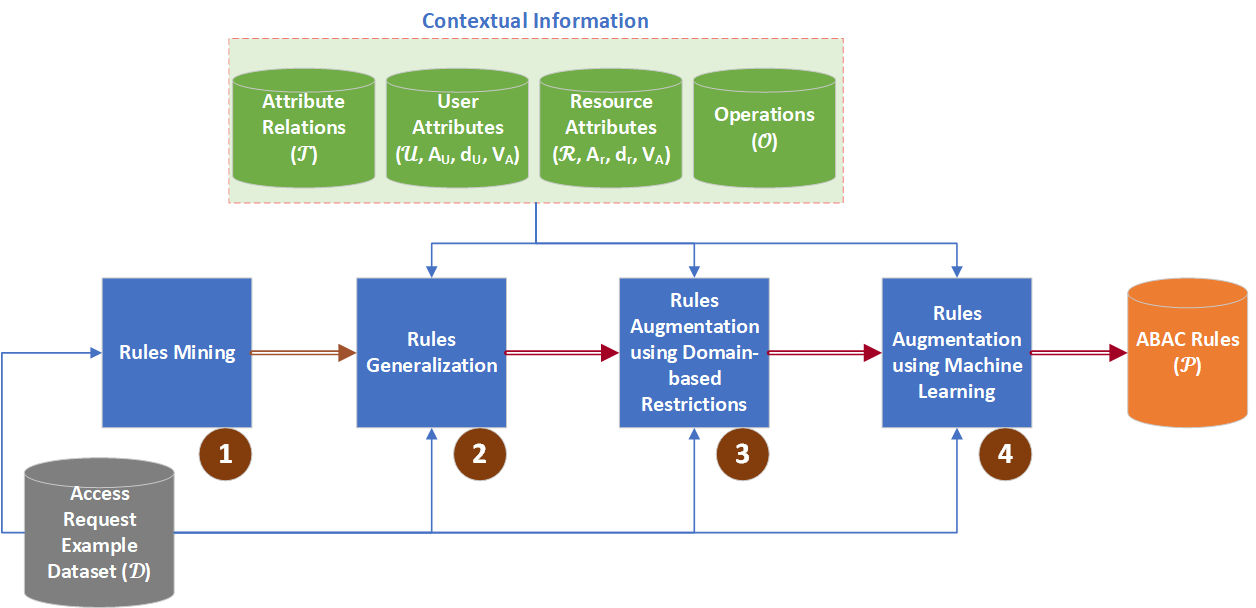}
    \caption{The Architecture of Polisma~\cite{Polisma}}
\label{fig:Polisma}
\end{figure*}

With the availability of a log of access requests and their corresponding control decisions, one can analyze such a log and generate the specifications of access control policies to control future requests. This problem is referred to as policy learning (a.k.a. policy mining) and several approaches have been proposed to solve this problem~\cite{xu2014mining}~\cite{cotrini2018mining}~\cite{sanders2019mining}. Recently, we introduced Polisma~\cite{Polisma}, a novel framework which utilizes both examples of access requests and corresponding access control decisions as well as contextual information obtained from other data sources, such as organizational charts,  user directories (e.g., directories of Lightweight Directory Access Protocol (LDAP),  workflows, security tracking's logs (e.g., logs provided by security information and event management systems), if available.

In Polisma (see Figure~\ref{fig:Polisma}), the learning process is performed according to a sequence of steps. A data mining technique is first used to infer associations between parties and resources from the set of decision examples, and based on these associations a set of rules is generated. In the second step, each constructed rule is generalized based on statistically significant attributes and context information. In the third step, policy domains are analyzed to augment the rules with restrictions as for some application domains (e.g., security) generalization can have undesired consequences. Policies learned by those stages are safe generalizations with minimal overfitting. To improve the completeness of the learned set, Polisma applies a ML classifier to decision examples not covered by the learned rules; it uses the classification result to generate additional rules in an ``ad-hoc'' manner.

\subsection{Problem Definition}

Policy learning algorithms typically analyze past access requests associated with their control decisions (see Definition~\ref{def:DE}) and aims at generating a set of ABAC rules to control any access request.

\begin{definition}[Access Control Decisions and Examples ($l$)] 
\label{def:DE}
An access control decision is a tuple $\langle u, r, o, d \rangle$ where $u$ is the user who initiated the access request, $r$ is the resource target of the access request, $o$ is the operation requested on the resource, and $d$ is the decision taken for the access request. A decision example $l$ is a tuple $\langle u,e_U,r,e_R,o,d \rangle$ where $e_U$ and $e_R$ are a user attribute expression, and a resource attribute expression such that $u$ $\models$ $e_U$, and $r$ $\models$ $e_R$, and the other arguments are interpreted as in an access control decision.
\end{definition}

In a coalition environment, different parties are involved and each party might encounter different scenarios that enrich their experience with respect to control access requests. However, such expertise might not be similar for all parties. These differences open the opportunity for collaboration among the parties and sharing of their experience and access control decisions. Therefore, in this work we assume that there are two parties willing to exchange knowledge about access control decisions. In particular, one party, called source party, has a set of ABAC policies (referred to source ABAC policies) and another party, called target party, has a set of access control decision examples, extracted from a system history of access requests and their corresponding authorized/denied decisions, together with some context information. The problem is to generate ``local'' ABAC policies at the target party utilizing both a local log as well as the ABAC policies from the source party. More precisely:

\begin{definition}[Transferring ABAC Policies by Local Examples and Context ($TAPEC$)]
\label{def:TAPEC}
Let $S$ and $T$ be the source party and $T$ be the target party. Assume that he following information is given:

\begin{itemize}
    \item A set of local access control decision examples (i.e., $\mathcal{D} = \{l_1, l_2, \dots, l_n\}$) at $T$.
    \item Local context information at $T$; namely, the sets $\mathcal{U}$, $\mathcal{R}$, $\mathcal{O}$, the sets $A_U$ and $A_R$, one of which could be empty, and the functions assigning values to the attributes of the users and resources. 
    \item A set of ABAC rules (i.e., $\mathcal{P}_S = \{{\rho}_{1} \dots {\rho}_{m}\}$) implemented or learned at $S$. 
\end{itemize}
$TAPEC$ aims at generating a set of ABAC rules (i.e., $\mathcal{P}_L = \{{\rho}_{1} \dots {\rho}_{w}\}$) that are able to control access requests at $T$.
\end{definition}
\section{Methodology}
\label{sec:approaches}
Transferring ABAC policies from a party to another includes three operations. First, the rules from the source party are compared with access request decision examples and the generated rules at the target party. Second, some rules potentially require adaptation to accommodate the differences between the domains and  context of the target party. Thus an adaptation process has to be executed. Finally, the original rules (from the source party) along with the new rules generated from the adaptation process, as well as the context information of the target party, are incorporated to enrich the target party with the security requirements from the source party.

\subsection{Rule Similarity Analysis}
The first step for transferring policies from a source party to a target party is to assess the similarity of each source rule with respect to the log of access control decision examples available at the target party. This process includes verifying that an access control decision example satisfies a rule of interest. Specifically, the verification comprises checking three conditions (see Definition~\ref{def:rule_example_similarity}): a) the user of the decision example satisfies the user attribute expression of the rule; b) the resource of the decision example satisfies the resource attribute expression of the rule; and c) the operation of the decision example is included in the rule operations set. Moreover, the target party might generate a set of rules (as we discuss
later on the paper). In this case, the policy transfer procedure requires checking the similarity between the source and target rules (see Definition~\ref{def:rules_similarity}).

Recognizing rule similarity enables next checking rule consistency. A rule is consistent with another similar rule if their corresponding decisions are identical (i.e., both of them are \textit{deny} or \textit{permit}); otherwise they are inconsistent. Similarly, an access control decision example is consistent with a similar rule if their decisions are identical. The consistency of a rule with respect to a decision example or another rule is formally defined in Definitions~\ref{def:rule_example_consistency} and~\ref{def:rules_consistency}, respectively.

\begin{definition}[Similarity of ABAC Rule and Access Control Decision Example] 
\label{def:rule_example_similarity}
Given an ABAC rule ${\rho} = \langle e_U, e_R, O, d \rangle$ and an access control decision example $l =\langle u,e_U,r,e_R,o,d \rangle$, $l$ is similar to ${\rho}$ (i.e., $l \approx {\rho}$) if and only if:
\begin{itemize}
\item ($l.u \models \rho.e_U \lor l.e_U \subseteq \rho.e_U$),
\item ($l.r \models \rho.e_R \lor l.e_R \subseteq \rho.e_R$), and
\item $l.o \subseteq \rho.O$.
\end{itemize}
\end{definition}

\begin{definition}[Consistency of ABAC Rule and Access Control Decision Example] 
\label{def:rule_example_consistency}
An ABAC rule ${\rho}$ and access control decision example $l$ are consistent (i.e., ${\rho} \simeq l$) if and only if:
\begin{itemize}
\item ${\rho} \approx l$, and
\item ${\rho}.d = l.d$.
\end{itemize}
\end{definition}

\begin{definition}[Similarity of a Pair of ABAC Rules] 
\label{def:rules_similarity}
Given two ABAC rules ${\rho}_i = \langle e_U, e_R, O, d \rangle$ and ${\rho}_j =\langle e_U,e_R,O,d \rangle$, ${\rho}_i$ and ${\rho}_j$ are similar to each other (i.e., ${\rho}_i \approx {\rho}_j$) if and only if:
\begin{itemize}
\item $({\rho}_i.e_U \subseteq {\rho}_j.e_U \lor {\rho}_i.u \models {\rho}_j.e_U)$ $\lor$ $({\rho}_j.e_U \subseteq {\rho}_i.e_U \lor {\rho}_j.u \models {\rho}_i.e_U)$,
\item $({\rho}_i.e_R \subseteq {\rho}_j.e_R \lor {\rho}_i.r \models {\rho}_j.e_R)$ $\lor$ $({\rho}_j.e_R \subseteq {\rho}_i.e_R \lor {\rho}_j.r \models {\rho}_i.e_R)$, and 
\item $({\rho}_i.o \subseteq {\rho}_j.O)$ $\lor$ $({\rho}_j.o \subseteq {\rho}_i.O)$.
\end{itemize}
\end{definition}

\begin{definition}[Consistency of a Pair of ABAC Rules] 
\label{def:rules_consistency}
Two rules ${\rho}_i$ and ${\rho}_j$ are consistent (i.e., ${\rho}_i \simeq {\rho}_j$) if and only if:
\begin{itemize}
\item ${\rho}_i \approx {\rho}_j$, and
\item ${\rho}_i.d = {\rho}_j.d$.
\end{itemize}
\end{definition}

\subsection{Rules Adaptation}
The second operation for policy transfer is to adapt the rules that are identified as inconsistent after conducting the similarity analysis. A straightforward approach to resolve the inconsistency between two rules (or a rule and an access request decision example) is to apply one of the policy-combining-algorithms used in XACML (e.g., deny overrides, permit overrides, and first applicable). However, prioritizing one of the rules on the other and ignoring the one with the least priority may lead to unintended consequences that increase over-privileged or under-privileged accesses. Thus, to avoid such scenarios it is safer to adapt the inconsistent rules in a way that resolves the inconsistency and preserves the intended privilege at the same time.

\subsubsection{Adaptation of Two Inconsistent Rules}
Resolving the conflict of two inconsistent rules (${\rho}_i$, ${\rho}_j$) is performed by first identifying the mutual and non-mutual rule predicates (i.e., operation, user, and resource predicates) between the two rules (see Definition~\ref{def:Mutual_Unmutual_User}), and then deriving new rules from the original ones based on either the identified mutual or non-mutual conditions (see Algorithm~\ref{alg:AdaptTwoRules}). Therefore, the first derived rule is generated based on the mutual rule predicates while setting the access control decision as either always ``permit'' (i.e., permissive paradigm) or always ``deny'' (restrictive paradigm) (see lines 2-3 in Algorithm~\ref{alg:AdaptTwoRules}). This strategy is analogous to the XACML combining algorithms applied when encountering a conflict. Using either the permissive or restrictive paradigm is heuristically decided based on the number of conflicting decision examples that are resolved by each paradigm (see lines 4-10 in Algorithm~\ref{alg:AdaptTwoRules}). Next, new rules are generated by adapting the original rules using the non-mutual rule predicates. In this case, the adaptation is performed by employing the following mechanisms on both the inconsistent rules. 
\begin{itemize}
    \item \textit{user-based rule adaptation}: preserve the decision of the access request for only the subset of non-mutual users while the rule resource expression and operations remain the same (see lines 12-13 in Algorithm~\ref{alg:AdaptTwoRules}); 
    \item \textit{resource-based rule adaptation}: preserve the decision of the access request for only the subset of non-mutual resources while the rule user expression and operations remain the same (see lines 15-16 in Algorithm~\ref{alg:AdaptTwoRules}); and 
    \item \textit{operation-based rule adaptation}: preserve the decision of the access request for only the subset of non-mutual operations while the rule user and  resource expressions remain the same (see lines 18-19 in Algorithm~\ref{alg:AdaptTwoRules}).
\end{itemize}
Nonetheless, employing only one of these mechanisms may result in the loss of some authorizations which are specified in the original rules. Thus, all of these mechanisms are applied in sequence. Nonetheless, to perform each of the adaptation mechanisms successfully, the corresponding non-mutual predicate should be a non-empty predicate (i.e., $U_n \neq \phi \lor R_n \neq \phi \lor O_n \neq \phi$). 
Finally, the original inconsistent rules are overridden by the rules derived from both of mutual and non-mutual rule predicates.

\begin{definition}[Mutual Predicates of ABAC Rules] 
\label{def:Mutual_Unmutual_User}
Given two rules ${\rho}_i = \langle e_U, e_R, O, d \rangle$ and ${\rho}_j = \langle e_U, e_R, O, d \rangle$, the mutual predicates comprise:
\begin{itemize}
\item the mutual user expression $U_m$: the intersection between the user expressions of both rules (i.e., $U_m = \{{\rho}_i.e_U \cap {\rho}_j.e_U\}$).
\item the mutual resource expression $R_m$: the intersection between the resource expressions of both rules  (i.e., $R_m = \{{\rho}_i.e_R \cap {\rho}_j.e_R\}$).
\item the mutual operations $O_m$: the subset of operations that are part of the operations of both rules (i.e., $O_m = \{{\rho}_i.O \cap {\rho}_j.O\}$).
\end{itemize}
Such that ($U_m$ $\neq$ $\phi$) $\land$ ($R_m$ $\neq$ $\phi$) $\land$ ($O_m$ $\neq$ $\phi$).
\end{definition}

\begin{definition}[Non-Mutual Predicates of ABAC Rules] 
\label{def:Mutual_Unmutual_User}
Given two rules ${\rho}_i = \langle e_U, e_R, O, d \rangle$ and ${\rho}_j = \langle e_U, e_R, O, d \rangle$, the non-mutual predicates comprise:
\begin{itemize}
\item the non-mutual user expression $U_n$: the user expression that is a subset of the user expression of ``only one'' of the two rules (i.e., $U_n$ = \{(${\rho}_i.e_U$ $\setminus$ ${\rho}_j.e_U$) $\lor$ (${\rho}_j.e_U$ $\setminus$ ${\rho}_i.e_U$)\}.
\item the non-mutual resource expression $U_n$: the resource expression that is a subset of the resource expression of ``only one'' of the two rules (i.e., $R_n = \{({\rho}_i.e_R \setminus {\rho}_j.e_R) \lor ({\rho}_j.e_R \setminus {\rho}_i.e_R$)\}.
\item the non-mutual operations $O_n$: the subset of operations are part of the operations of ``only one'' of the two rules (i.e., $O_n = \{({\rho}_i.O \setminus {\rho}_j.O) \lor ({\rho}_j.O \setminus {\rho}_i.O$)\}.
\end{itemize}
\end{definition}

\begin{algorithm}[!htb]
\caption{Adaptation of Two Inconsistent ABAC Rules ($Adapt2Rules$)}
\label{alg:AdaptTwoRules}
\begin{algorithmic}[1]
\REQUIRE $\mathcal{D}$: Access control decision examples and (${\rho}_i$, ${\rho}_j$) a pair of inconsistent ABAC rules 
\STATE $\mathcal{P}$ = \{\}
\STATE ${\rho}_k$ = $\langle$ ${\rho}_i.e_U$ $\cap$ ${\rho}_j.e_U$, ${\rho}_i.e_R$ $\cap$ ${\rho}_j.e_R$, ${\rho}_i.o$ $\cap$ ${\rho}_j.o$, $permit$ $\rangle$
\STATE ${\rho}_k^{\prime}$ = $\langle$ ${\rho}_i.e_U$ $\cap$ ${\rho}_j.e_U$, ${\rho}_i.e_R$ $\cap$ ${\rho}_j.e_R$, ${\rho}_i.o$ $\cap$ ${\rho}_j.o$, $deny$ $\rangle$
\STATE $\mathcal{D}_{{\rho}_{k}}$ = \{$\forall l$ $\in$ $\mathcal{D}$ $\mid$ (${l} \simeq {\rho}_k$ \}
\STATE $\mathcal{D}_{{\rho}_{k}^{\prime}}$ = \{$\forall l$ $\in$ $\mathcal{D}$ $\mid$ (${l} \simeq {\rho}_{k}^{\prime}$ \}
\IF {$\mid\mathcal{D}_{{\rho}_{k}}\mid$ $>$ $\mid\mathcal{D}_{{\rho}_{k}^{\prime}}\mid$}
\STATE $\mathcal{P} = \mathcal{P} \cup {\rho}_{k}$
\ELSE
\STATE $\mathcal{P} = \mathcal{P} \cup {\rho}_{k\prime}$
\ENDIF
\STATE ${\rho}_i^{\prime}$ = $\langle$ ${\rho}_i.e_U$ $\setminus$ ${\rho}_j.e_U$, ${\rho}_i.e_R$, ${\rho}_i.O$, ${\rho}_i.d$ $\rangle$
\STATE ${\rho}_j^{\prime}$ = $\langle$ ${\rho}_j.e_U$ $\setminus$ ${\rho}_i.e_U$, ${\rho}_j.e_R$, ${\rho}_j.O$, ${\rho}_j.d$ $\rangle$
\STATE ${\rho}_i^{\prime\prime}$ = $\langle$ ${\rho}_i.e_U$, ${\rho}_i.e_R$ $\setminus$ ${\rho}_j.e_R$, ${\rho}_i.O$, ${\rho}_i.d$ $\rangle$
\STATE ${\rho}_j^{\prime\prime}$ = $\langle$ ${\rho}_j.e_U$, ${\rho}_j.e_R$ $\setminus$ ${\rho}_i.e_R$, ${\rho}_j.O$, ${\rho}_j.d$ $\rangle$
\STATE ${\rho}_i^{\prime\prime\prime}$ = $\langle$ ${\rho}_i.e_U$, ${\rho}_i.e_R$, ${\rho}_i.O$ $\setminus$ ${\rho}_j.O$, ${\rho}_i.d$ $\rangle$
\STATE ${\rho}_j^{\prime\prime\prime}$ = $\langle$ ${\rho}_j.e_U$, ${\rho}_j.e_R$, ${\rho}_j.O$ $\setminus$ ${\rho}_i.O$, ${\rho}_j.d$ $\rangle$
\STATE $\mathcal{P} = \mathcal{P} \cup {\rho}_i^{\prime} \cup {\rho}_j^{\prime} \cup {\rho}_i^{\prime\prime} \cup {\rho}_j^{\prime\prime} \cup {\rho}_i^{\prime\prime\prime} \cup {\rho}_j^{\prime\prime\prime}$
\RETURN $\mathcal{P}$
\end{algorithmic}
\end{algorithm}

\textit{Example: Given two inconsistent rules
${\rho}_1$ = $\langle$ \{dept id: 9, 10, 11\}, \{resource id: 1, 2, 3\}, \{read, write\}, permit $\rangle$, 
${\rho}_2$ = $\langle$ \{dept id: 9, 12\}, \{resource id: 1\}, \{read\}, deny $\rangle$, 
the resolution is performed as follows:\\
For the non-mutual predicates:
By the user-based adaptation, the new rules are:
        \begin{itemize}
            \item ${\rho}_1^{\prime}$=$\langle$ \{dept id: 10,11\}, \{resource id: 1,2,3\}, \{read,write\}, permit $\rangle$
            \item ${\rho}_2^{\prime}$=$\langle$ \{dept id: 12\}, \{resource id: 1\}, \{read\}, deny $\rangle$
        \end{itemize}
By the resource-based adaptation, the new rules are:
        \begin{itemize}
            \item ${\rho}_1^{\prime\prime}$=$\langle$ \{dept id: 9,10,11\}, \{resource id: 2,3\}, \{read,write\}, permit $\rangle$
        \end{itemize}
By the operation-based adaptation, the new rules are:
        \begin{itemize}
            \item ${\rho}_1^{\prime\prime\prime}$=$\langle$ \{dept id: 9,10,11\}, \{resource id: 1,2,3\}, \{write\}, permit $\rangle$.
        \end{itemize}
For the mutual predicates:
\begin{itemize}
\item According to the permissive paradigm, the new rule is : ${\rho}_2^{\prime\prime}$ = $\langle$ \{dept id: 9\}, \{resource id: 1\}, \{read\}, permit $\rangle$
\item According the restrictive paradigm, the new rule is : ${\rho}_2^{\prime\prime\prime}$ = $\langle$ \{dept id: 9\}, \{resource id: 1\}, \{read\}, deny $\rangle$
 \end{itemize}
}

Eventually, when adapting two inconsistent rules, at maximum six rules can be generated based on the non-mutual predicates, while one rule at maximum can be generated based on the mutual predicates.

\subsubsection{Adaptation of Groups of Inconsistent Rules}
When an ABAC rule contradicts a set of ABAC rules, one approach is to execute the resolution algorithm separately for each pair of inconsistent rules (i.e., Algorithm~\ref{alg:AdaptTwoRules}). However, this might not be as simple as it seems. Applying such an approach potentially generates additional conflicts because the newly adapted rules are generated by considering only a pair of rules but not the other ones. Thus, a recursive adaptation is performed until no further inconsistency is encountered within the original group of inconsistent rules and the ones adapted from them (i.e., the newly generated ones).   

\begin{algorithm}[!htb]
\begin{algorithmic}[1]
\caption{Subtract Rules ($SubAttrFromRule$)}
\label{alg:substractAttributes}
\REQUIRE An ABAC rule ${\rho}_0$ inconsistent with a group of rules ($\mathcal{P}_\mathcal{C}$ = \{${\rho}_1$, $\dots$, ${\rho}_m$\}) and $\mathcal{P}$ the set of resulted rules 
\IF{$\mid\mathcal{P}_{\mathcal{C}}\mid = 0$}
\RETURN ${\rho}_0$
\ELSE
\FOR{${a}_{i} \in ({\rho}_0.a_U \cup {\rho}_0.a_R \cup {\rho}_0.O$)}
\STATE $\rho_1 = \mathcal{P}_{\mathcal{C}} \setminus \{\forall {\rho}_i \in \mathcal{P}_{\mathcal{C}} \mid i > 1 \}$
\IF {${a}_{i} \in {\rho}_0.a_U$}
\STATE ${e_U}_{a0} = \{(a_i,d_U(u, a_i))\mid \forall u \models {\rho}_0.a_U\}$
\STATE ${e_U}_{a1} = \{(a_i,d_U(u, a_i))\mid \forall u \models {\rho}_1.a_U\}$
\STATE $a_U^{\prime} = \{\forall u \models {e_U}_{a0} \} \setminus \{\forall u \models {e_U}_{a1}\}$
\STATE ${\rho}_0^{\prime} = \langle a_U^{\prime}, \rho_0.a_R, \rho_0.O, \rho_0.d \rangle$
\ELSIF{${a}_{i} \in {\rho}_0.a_R$}
\STATE ${e_R}_{a0} = \{(a_i,d_R(r, a_i))\mid \forall r \models {\rho}_0.a_R\}$
\STATE ${e_R}_{a1} = \{(a_i,d_R(r, a_i))\mid \forall r \models {\rho}_1.a_R\}$ 
\STATE $a_R^{\prime} = \{\forall r \models {e_R}_{a0} \} \setminus \{\forall r \models {e_R}_{a1}\}$
\STATE ${\rho}_0^{\prime} = \langle \rho_0.a_U, a_R^{\prime}, \rho_0.O, \rho_0.d \rangle$
\ELSIF{${a}_{i} \in {\rho}_0.O$}
\STATE $a_O^{\prime} = {\rho}_0.O \setminus {\rho}_1.O$
\STATE ${\rho}_0^{\prime} = \langle \rho_0.a_U, \rho_0.a_R, a_O^{\prime}, \rho_0.d \rangle$
\ENDIF
\IF{${\rho}_0^{\prime} \neq \phi$} 
\STATE ${\rho}_0 = {\rho}_0^{\prime}$ 
\STATE $\mathcal{P}_{\mathcal{C}} = \{\forall {\rho}_i \in \mathcal{P}_{\mathcal{C}} \mid i > 1 \}$
\STATE $\mathcal{P}$ = $\mathcal{P}$ $\cup$ $SubAttrFromRule$(${\rho}_0, \mathcal{P}_{\mathcal{C}}, \mathcal{P}$)
\ENDIF
\ENDFOR
\ENDIF
\RETURN $\mathcal{P}$
\end{algorithmic}
\end{algorithm}

\begin{algorithm}[!htb]
\caption{Adaptation of a Group of Inconsistent ABAC Rules ($AdaptGRules$)}
\label{alg:AdaptMultiRules}
\begin{algorithmic}[1]
\REQUIRE $\mathcal{D}$: Access control decision examples and an ABAC rule ${\rho}_0$ inconsistent with a group of rules ($\mathcal{P}_\mathcal{G}$ = \{${\rho}_1$, $\dots$, ${\rho}_m$\}) 
\STATE $\mathcal{P}_{\mathcal{C}}$ = \{\}, $\mathcal{P}$ = \{\}
\FOR{${\rho}_{i} \in \mathcal{P}_{\mathcal{G}}$}
\STATE $\mathcal{P}_{\mathcal{C}} = \mathcal{P}_C \cup ({\rho}_{0} \cap {\rho}_{i})$
\ENDFOR
\STATE $\mathcal{P}$  = $SubAttrFromRule$(${\rho}_0, \mathcal{P}_{\mathcal{C}}, \mathcal{P}$)
\FOR{${\rho}_{i} \in \mathcal{P}_{\mathcal{G}}$}
\STATE $\mathcal{P} = \mathcal{P} \cup SubAttrFromRule({\rho}_{i}, \{{\rho}_0 \cap {\rho}_{i}\}, \mathcal{P})$
\ENDFOR
\FOR{${\rho}_{i} \in \mathcal{P}_{\mathcal{C}}$}
\STATE ${\rho}_k$ = $\langle$ ${\rho}_i.e_U$, ${\rho}_i.e_R$, ${\rho}_i.o$, $permit$ $\rangle$
\STATE ${\rho}_k^{\prime}$ = $\langle$ ${\rho}_i.e_U$, ${\rho}_i.e_R$, ${\rho}_i.o$, $deny$ $\rangle$
\STATE $\mathcal{D}_{{\rho}_{k}}$ = \{$\forall l$ $\in$ $\mathcal{D}$ $\mid$ (${l} \simeq {\rho}_k$ \}
\STATE $\mathcal{D}_{{\rho}_{k}^{\prime}}$ = \{$\forall l$ $\in$ $\mathcal{D}$ $\mid$ (${l} \simeq {\rho}_{k}^{\prime}$ \}
\IF {$\mid\mathcal{D}_{{\rho}_{k}}\mid$ $>$ $\mid\mathcal{D}_{{\rho}_{k}^{\prime}}\mid$}
\STATE $\mathcal{P} = \mathcal{P} \cup {\rho}_{k}$
\ELSE
\STATE $\mathcal{P} = \mathcal{P} \cup {\rho}_{k\prime}$
\ENDIF
\ENDFOR
\RETURN $\mathcal{P}$
\end{algorithmic}
\end{algorithm}

\subsection{Rule Transferability Approaches}
In what follows, we first introduce two na\"ive approaches for rule transfer and then introduce our proposed approaches.
\label{subsec:RuleTransfer}
\subsubsection{Na\"ive Approaches}

\subsubsection*{Policies Transfer using a Local Log ({\em TPLG}).}

\begin{figure}[!htbp]
	\centering
    \includegraphics[width=0.9\columnwidth]{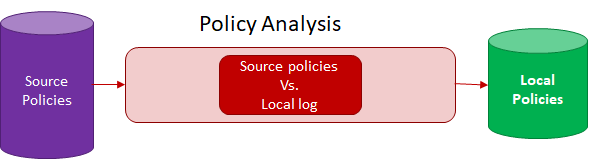}
    \caption{Policies Transfer using Local Logs ($TPLG$)}
\label{fig:TPLG}
\end{figure}

A straightforward approach for transferring policies is to use the raw log of historical access control decisions collected at the target party for adapting the source rules (see Figure~\ref{fig:TPLG}). In particular, the rules are tuned using the local log, according to Algorithm~\ref{alg:TPLG}, Towards this, first the source rules are enforced using the historical set of access control requests available. After the enforcement, the decision of each rule for the corresponding examined access requests is compared to the historical access control decisions. Such an inspection can result in three cases: a) the rule has not been enforced for any of the historical requests, b) the rule has been enforced with no inconsistency for any of the historical decisions, c) the rule has been enforced with inconsistency with respect to the historical decisions. In the first case, the rule is transferred in order to be used for handling situations not yet encountered. For the second case, the rule is also transferred because of its compliance with the local historical access control decisions. Finally, in the third case, the rule is adapted by restricting its attribute expressions to fit with either only the corresponding historical decision requests that comply with or the contradicting set of historical decision requests.

\begin{algorithm}[!htb]
\caption{Policies Transfer using Local Log ($TPLG$)}
\label{alg:TPLG}
\begin{algorithmic}[1]
\REQUIRE $\mathcal{D}$: Access control decision examples, $\mathcal{P}_{S}$: ``Source'' access control policies.
\STATE $\mathcal{P}_{L}$ = \{\}
\FOR{${\rho}_{i} \in \mathcal{P}_{S}$}
\STATE $\mathcal{D}_{{\rho}_{i}}$ = \{$\forall l$ $\in$ $\mathcal{D}$ $\mid$ $l$ $\approx$ ${\rho}_{i}$\}
\STATE $\mathcal{D}_{{\rho}_{i}}^{m}$ = $\{\forall l \in \mathcal{D}_{{\rho}_{i}} \mid l \simeq {\rho}_{i} \}$
\STATE $\mathcal{D}_{{\rho}_{i}}^{c}$ = $\{\forall l \in \mathcal{D}_{{\rho}_{i}} \mid l \not\simeq {\rho}_{i} \}$
\IF{$\mathcal{D}_{{\rho}_{i}}^{m} \neq \phi$ $\land$ $\mathcal{D}_{{\rho}_{i}}^{c} = \phi$}
\STATE $\mathcal{P}_{L} = \mathcal{P}_{L} \cup {\rho}_{i}$
\ELSIF{$\mathcal{D}_{{\rho}_{i}}^{c} \neq \phi$}
\STATE $\mathcal{P}_{{\rho}_{i}}^{c^{\prime}}$ = \{\}
\STATE $\mathcal{P}_{{\rho}_{i}}^{c^{\prime}}$ = $AdaptGRules$($\mathcal{D}$, ${\rho}_{i}$, $\mathcal{D}_{{\rho}_{i}}^{c^{\prime}}$)
\STATE $\mathcal{P}_{L} = \mathcal{P}_{L} \cup \mathcal{P}_{{\rho}_{i}}^{c^{\prime}}$
\ELSIF{$\mathcal{D}_{{\rho}_{i}}^{m} = \phi$ $\land$ $\mathcal{D}_{{\rho}_{i}}^{c} = \phi$}
\STATE $\mathcal{P}_{L} = \mathcal{P}_{L} \cup {\rho}_{i}$
\ENDIF
\ENDFOR
\RETURN $\mathcal{P}_{L}$
\end{algorithmic}
\end{algorithm}

\subsubsection*{Policies Transfer using Local Policies ({\em TPLP}).} 
$TPLG$ is easy to implement; however, its main limitation is that it utilizes the raw local log for tuning the source policies directly without analyzing the log. Such an approach assumes that the observation of each historical access control decision represents a rule; however, each rule is typically an abstraction of security specifications for multiple access control decisions. An alternative approach is to utilize the local log for generating ABAC rules referred to as ``local policies'' using one of the state-of-the-art policy learning approaches~\cite{Polisma}~\cite{xu2014mining}~\cite{cotrini2018mining}.

\begin{figure}[!htbp]
	\centering
    \includegraphics[width=1.0\columnwidth]{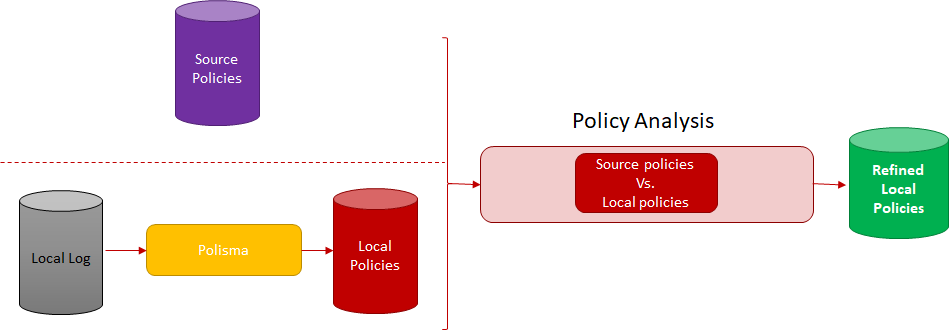}
    \caption{Policies Transfer using Local Policies ($TPLP$)}
\label{fig:TPLP}
\end{figure}

Such an approach is composed of two main steps (see Algorithm~\ref{alg:TPLP} and Figure~\ref{fig:TPLP}). First, the approach uses a policy learner to generate local rules using the local log. Thereafter, a similarity analysis is performed on both local and source rules. The outcome of such analysis has three cases: a) no local rule is similar to a source rule (see lines 11-12);  b) all the local rules, which are similar to a source rule, are consistent with it (see lines 6-7); and c) some of the local rules, which are similar to a source rule, are inconsistent with it (see lines 8-10). In the first and the second cases, the source rule is migrated to the target system. In the third case, the conflicts between the local and source rules are resolved by performing the adaptation algorithm such that their corresponding attribute expressions are tailored either by expanding or restricting their covered scope.

\begin{algorithm}[!htb]
\caption{Policies Transfer using Local Policies ($TPLP$)}
\label{alg:TPLP}
\begin{algorithmic}[1]
\REQUIRE $\mathcal{P}_{S}$: ``Source'' access control policies, $\mathcal{P}_{L}$: ``Local'' access control policies.
\STATE $\mathcal{P}_{L}^{\prime}$ = \{\}
\FOR{${\rho}_{i} \in \mathcal{P}_{S}$}
\STATE $\mathcal{P}_{{\rho}_{i}}$ = \{$\forall p_j$ $\in$ $\mathcal{P}_{L}$ $\mid$ ${p}_{j}$ $\approx$ ${\rho}_{i}$ \}
\STATE $\mathcal{P}_{{\rho}_{i}}^{m}$ = $\{\forall p_k \in \mathcal{P}_{{\rho}_{i}} \mid p_k \simeq {\rho}_{i} \}$
\STATE $\mathcal{P}_{{\rho}_{i}}^{c}$ = $\{\forall p_k \in \mathcal{P}_{{\rho}_{i}} \mid p_k \not\simeq {\rho}_{i} \}$
\IF{$\mathcal{P}_{{\rho}_{i}}^{m} \neq \phi$ $\land$ $\mathcal{P}_{{\rho}_{i}}^{c} = \phi$}
\STATE $\mathcal{P}_{L}^{\prime} = \mathcal{P}_{L}^{\prime} \cup {\rho}_{i}$
\ELSIF{$\mathcal{P}_{{\rho}_{i}}^{c} \neq \phi$}
\STATE $\mathcal{P}_{{\rho}_{i}}^{c^{\prime}}$ = \{\}
\STATE $\mathcal{P}_{{\rho}_{i}}^{c^{\prime}}$ = $AdaptGRules$($\mathcal{D}$, ${\rho}_{i}$, $\mathcal{P}_{{\rho}_{i}}^{c}$)
\STATE $\mathcal{P}_{L}^{\prime} = \mathcal{P}_{L}^{\prime} \cup \mathcal{P}_{{\rho}_{i}}^{c^{\prime}}$
\ELSIF{$\mathcal{P}_{{\rho}_{i}}^{m} = \phi$ $\land$ $\mathcal{P}_{{\rho}_{i}}^{c} = \phi$}
\STATE $\mathcal{P}_{L}^{\prime} = \mathcal{P}_{L}^{\prime} \cup {\rho}_{i}$
\ENDIF
\ENDFOR
\RETURN $\mathcal{P}_{L}^{\prime}$
\end{algorithmic}
\end{algorithm}

\subsubsection{Proposed Approaches}

\begin{figure*}[!htbp]
\begin{minipage}[t]{0.49\linewidth}
	\centering
    \includegraphics[width=1.0\linewidth]{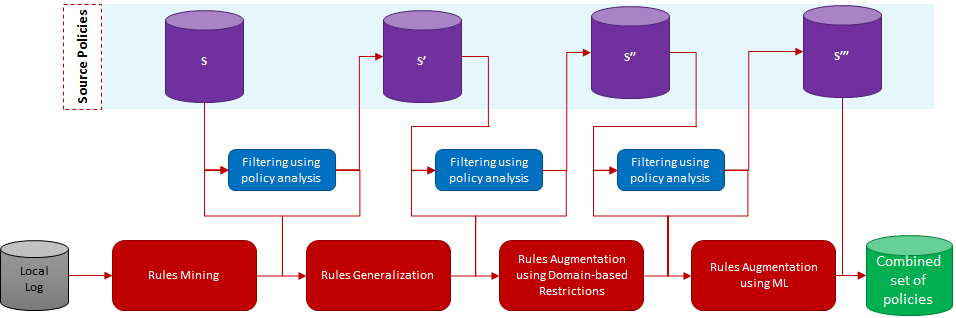}
    \caption{Policies Transfer using Local Learning }
    \label{fig:TPLL}
\end{minipage}
\begin{minipage}[t]{0.49\linewidth}
	\centering
    \includegraphics[width=1.0\linewidth]{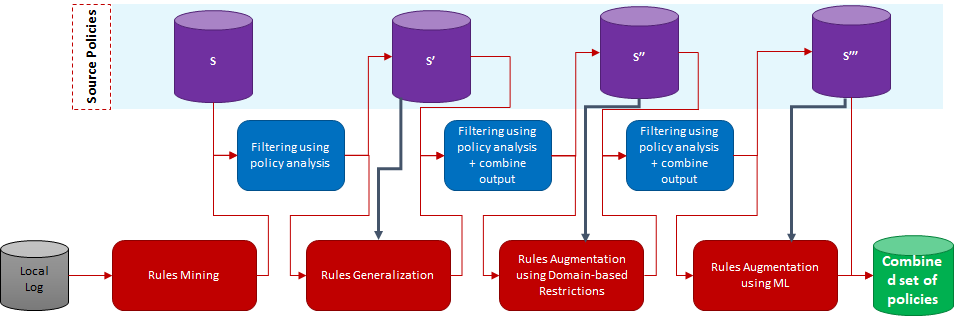}
    \caption{Policies Transfer using Hybrid Learning}
    \label{fig:TPHL}
\end{minipage}
\end{figure*}

\subsubsection*{Policies Transfer using Local Learning ({\em TPLL}.)} 

$TPLP$ post-processes the generated rules from a policy learner using the source rules. However, such a mechanism potentially generates local rules that are inconsistent with the source rules and also increases the conflicts among the rules learned throughout the learning stages. One approach to avoid such a problem is to use the source rules to adapt the intermediate forms of the local rules. In particular, the rule adaptation is performed in tandem with policy learning from the local log.

Since a policy learner is typically composed of multiple learning phases, the intermediate rules generated after each step are compared with the source rules using a similarity analysis (see Algorithm~\ref{alg:TPLL} and Figure~\ref{fig:TPLL}). If these intermediate local rules conflict with the source rules, the local rules are adapted in the early stages; hence enabling the evolution of more accurate local rules among the phases of the policy learner (i.e., reducing error propagation) while filtering the source policies to exclude the ones that have a conflict with any of the local rules. Moreover, this mechanism enables either increasing or decreasing the significance of intermediate rules and controlling their evolution. After the local rules are generated, the remaining source rules are transferred to the target system.

\begin{algorithm}[!htb]
\caption{Policies Transfer using Local Learning ($TPLL$)}
\label{alg:TPLL}
\begin{algorithmic}[1]
\REQUIRE $\mathcal{D}$: Access control decision examples, $\mathcal{P}_{S}$: ``Source'' access control policies, $\Uppsi$: a policy learner consisted of $n$ steps \{${\Upomega}_1$, ${\Upomega}_2$, $\dots$, ${\Upomega}_n$\}.
\STATE $\mathcal{P}_{S}^{\prime}$ = $\mathcal{P}_{S}$
\STATE $\mathcal{P}_{L}^{\prime}$ = \{\}
\FOR{${\Upomega}_{k} \in \Uppsi$}
\STATE $\mathcal{P}_{L}$ = ${\Upomega}_{k}$($\mathcal{D}$, $\mathcal{P}_{L}^{\prime}$)
\FOR{${\rho}_{i} \in \mathcal{P}_{S}^{\prime}$}
\STATE $\mathcal{P}_{{\rho}_{i}}$ = \{$\forall p_j$ $\in$ $\mathcal{P}_{L}$ $\mid$ ${p}_{j}$ $\approx$ ${\rho}_{i}$ \}
\STATE $\mathcal{P}_{{\rho}_{i}}^{m}$ = $\{\forall p_k \in \mathcal{P}_{{\rho}_{i}} \mid p_k \simeq {\rho}_{i} \}$
\STATE $\mathcal{P}_{{\rho}_{i}}^{c}$ = $\{\forall p_k \in \mathcal{P}_{{\rho}_{i}} \mid p_k \not\simeq {\rho}_{i} \}$
\IF{$\mathcal{P}^{{\rho}_{i}}_{c} \neq \phi$}
\STATE $\mathcal{P}_{{\rho}_{i}}^{c^{\prime}}$ = \{\}
\STATE $\mathcal{P}_{{\rho}_{i}}^{c^{\prime}}$ = $AdaptGRules$($\mathcal{D}$, ${\rho}_{i}$, $\mathcal{P}_{{\rho}_{i}}^{c}$)
\STATE $\mathcal{P}_{S}^{\prime} = \mathcal{P}_{S}^{\prime} - {\rho}_{i}$
\STATE $\mathcal{P}_{L} = \mathcal{P}_{L} - \mathcal{P}_{{\rho}_{i}}^{c}$
\STATE $\mathcal{P}_{L} = \mathcal{P}_{L} \cup \mathcal{P}_{{\rho}_{i}}^{c^{\prime}}$
\ENDIF
\ENDFOR
\STATE $\mathcal{P}_{L}^{\prime}$ = $\mathcal{P}_{L}$
\ENDFOR
\STATE $\mathcal{P}_{L}^{\prime} = \mathcal{P}_{L}^{\prime} \cup \mathcal{P}_{S}^{\prime}$
\RETURN $\mathcal{P}_{L}^{\prime}$
\end{algorithmic}
\end{algorithm}

\subsubsection*{Policies Transfer using Hybrid Learning ({\em TPHL}).}

$TPLL$ does not fully exploit the source rules while learning the local rules from the local log since $TPLL$ only uses the source rules for controlling the evolution of the intermediate rules. Therefore, we propose another approach in which the source rules are in-lined with the intermediate local rules to allow the next learning phases to exploit the source rules as well as the local log in the learning stages. Such an approach allows one to generate correct rules that cover further security aspects and requirements of the local system.

As shown in Figure~\ref{fig:TPHL}, the source rules are used in two ways in this approach. First, the intermediate local rules are adapted after each learning phase of the policy learner. Second, the filtered source rules (by excluding the ones that have a conflict with any of the local rules) are used as input for the subsequent learning phases alongside the intermediate local rules. The steps of $TPHL$ are illustrated in Algorithm~\ref{alg:TPLH}.

\begin{algorithm}[!htb]
\caption{Policies Transfer using Hybrid Learning ($TPHL$)}
\label{alg:TPLH}
\begin{algorithmic}[1]
\REQUIRE $\mathcal{D}$: Access control decision examples, $\mathcal{P}_{S}$: ``Source'' access control policies, $\Uppsi$: a policy learner consisted of $n$ steps \{${\Upomega}_1$, ${\Upomega}_2$, $\dots$, ${\Upomega}_n$\}.
\STATE $\mathcal{P}_{S}^{\prime}$ = $\mathcal{P}_{S}$.
\STATE $\mathcal{P}_{L}^{\prime}$ = \{\}.
\FOR{${\Upomega}_{k} \in \Uppsi$}
\STATE $\mathcal{P}_{L}$ = ${\Upomega}_{k}$($\mathcal{D}$, $\mathcal{P}_{L}^{\prime}$)
\FOR{${\rho}_{i} \in \mathcal{P}_{S}^{\prime}$}
\STATE $\mathcal{P}_{{\rho}_{i}}$ = \{$\forall p_j$ $\in$ $\mathcal{P}_{L}$ $\mid$ ${p}_{j}$ $\approx$ ${\rho}_{i}$ \}
\STATE $\mathcal{P}_{{\rho}_{i}}^{m}$ = $\{\forall p_k \in \mathcal{P}_{{\rho}_{i}} \mid p_k \simeq {\rho}_{i} \}$
\STATE $\mathcal{P}_{{\rho}_{i}}^{c}$ = $\{\forall p_k \in \mathcal{P}_{{\rho}_{i}} \mid p_k \not\simeq {\rho}_{i} \}$
\IF{$\mathcal{P}_{{\rho}_{i}}^{c} \neq \phi$}
\STATE $\mathcal{P}_{{\rho}_{i}}^{c^{\prime}}$ = \{\}
\STATE $\mathcal{P}_{{\rho}_{i}}^{c^{\prime}}$ = $AdaptGRules$($\mathcal{D}$, ${\rho}_{i}$, $\mathcal{P}_{{\rho}_{i}}^{c}$)
\STATE $\mathcal{P}_{S}^{\prime} = \mathcal{P}_{S}^{\prime} - {\rho}_{i}$
\STATE $\mathcal{P}_{L} = \mathcal{P}_{L} - {\rho}_{i}$
\STATE $\mathcal{P}_{L} = \mathcal{P}_{L} - \mathcal{P}^{{\rho}_{i}}_{c}$
\STATE $\mathcal{P}_{L} = \mathcal{P}_{L} \cup \mathcal{P}_{{\rho}_{i}}^{c^{\prime}}$
\ENDIF
\ENDFOR
\STATE $\mathcal{P}_{L} = \mathcal{P}_{L} \cup \mathcal{P}_{S}^{\prime}$
\STATE $\mathcal{P}_{L}^{\prime}$ = $\mathcal{P}_{L}$
\ENDFOR
\RETURN $\mathcal{P}_{L}^{\prime}$
\end{algorithmic}
\end{algorithm}
\section{Evaluation}
\label{sec:evaluation}

In this section, we summarize the experimental methodology and report the evaluation results for \textit{FLAP}.
\begin{figure*}[!htbp]
\begin{minipage}[t]{0.49\linewidth}
	\centering
    \includegraphics[width=1.0\linewidth]{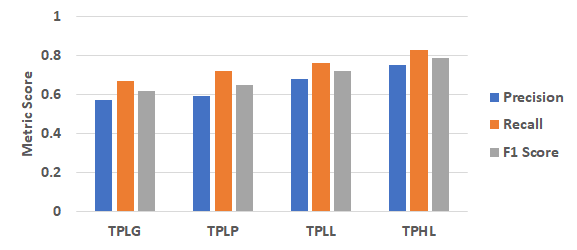}
    \caption{Policies Transfer using the $PM$ dataset }
    \label{fig:PM_Exp}
\end{minipage}
\begin{minipage}[t]{0.49\linewidth}
	\centering
    \includegraphics[width=1.0\linewidth]{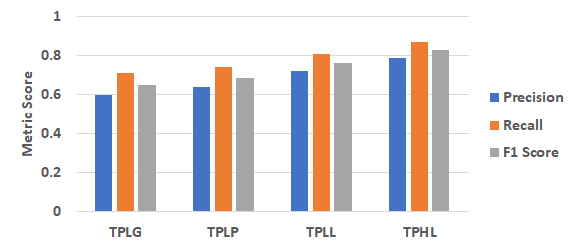}
    \caption{Policies Transfer using the Amazon dataset}
    \label{fig:AZ_Exp}
\end{minipage}
\vspace{-5mm}
\end{figure*}
\subsection{Experimental Methodology}

\noindent\textit{\textbf{Dataset.}} We conducted experiments using two datasets: a synthetic dataset (referred to as project management ($PM$) dataset~\cite{xu2014mining}) and real dataset (referred to as Amazon dataset), obtained from Amazon\footnote{\url{http://archive.ics.uci.edu/ml/datasets/Amazon+Access+Samples}}. This dataset is an anonymized sample of accesses provisioned within the Amazon company and it is composed of around 700K decision examples. We used 70\% of the dataset as decision examples of the target party and 30\% for the source party.\\
\noindent\textit{\textbf{Evaluation Metrics.}} To evaluate \textit{FLAP}, we use three metrics: precision, recall, and F1 score. These metrics are able to assess the correctness of the adapted policies, by checking their responses to the local decision examples, and their ability to cover new access requests.

\subsection{Experimental Results}
Figures~\ref{fig:PM_Exp} and \ref{fig:AZ_Exp} shows the evaluation results for the four approaches introduced in the paper, that is, $TPLG$, $TPLP$, $TPLL$, and $TPHL$, on the $PM$ and Amazon datasets, respectively. In general, both $TPLL$ and $TPHL$ perform better than $TPLG$, $TPLP$ in terms of all evaluation metrics due to the fact that both $TPLL$ and $TPHL$ utilize the source policies along with the local log in the learning process to filter and adapt the rules including the intermediate ones; hence reducing the effect of  error propagation. On the other hand, $TPLG$ and $TPLP$ perform the worst because $TPLG$ uses ``only'' the local log for adapting the source policies without analyzing the local log while $TPLP$ uses the local log to learn local rules for the target system without utilizing the source rules. Moreover, the results on the Amazon dataset are better than that of the $PM$ dataset because the Amazon dataset contains more decision examples in the local log and more source policies compared to the $PM$ dataset, and thus, such an increase in the input data positively affects the transfer process. Figures~\ref{fig:PM_Exp} and \ref{fig:AZ_Exp} show the evaluation results with respect to three metrics: precision, recall, and F1 score. The recall metric reflects the ability of the rules generated of the transfer process to correctly control more scenarios, while the precision metric reflects the correctness of the decisions produced by the generated rules\footnote{F1 score is the harmonic mean of precision and recall.}.
\section{Related Work}
\label{sec:relatedwork}
The problem of ABAC policy transfer has not yet been investigated. Approaches have been proposed for policy adaptation in the context of mobile systems, such as~\cite{Efstratiou}. However the policies considered in such approaches are not attribute-based. Work related to ours also includes work on mining ABAC policies from access control decision logs~\cite{cotrini2018mining},~\cite{ Polisma},~\cite{sanders2019mining},~\cite{mocanu2015towards},~\cite{xu2014mining}, and work on ABAC policy similarity~\cite{lin2010exam},~\cite{lin2007approach}. However such approaches do not address policy migration and adaptation. More recently approaches have been proposed that apply transfer learning techniques in the context of intrusion detection~\cite{ankush2020},~\cite{ankush2019},~\cite{zhao2019transfer},~\cite{zhao2017feature},~\cite{tzeng2017adversarial}. However those approaches have been designed for transferring neural networks used for classifying network packets as benign or malicious and thus do not deal with rule transfer and adaptation. 
\section{Conclusion and Future Work}
\label{sec:conclusion}
In this paper, we have proposed \textit{FLAP}, a framework for policy transfer in a federated environments. It allows one to transfer attribute-based policies from a source party to a target party. We have proposed four approaches that vary in their interaction levels between the source and target resources, as well as the timing of this interaction. Our preliminary evaluation, carried out on a real-world access control decision dataset and a synthetic one, show the ability of \textit{FLAP} to transfer the source policies to a target party to generate correct policies which can be used in future for unseen scenarios. As part of future work, we plan to perform experiments to assess the effect of different factors and scenarios in the transfer process, including data sharing percentage and types.

\section{Acknowledgments}
This research was sponsored by the U.S. Army Research Laboratory and the U.K. Ministry of Defence under Agreement Number W911NF-16-3-0001. The views and conclusions contained in this document are those of the authors and should not be interpreted as representing the official policies, either expressed or implied, of the U.S. Army Research Laboratory, the U.S. Government, the U.K. Ministry of Defence or the U.K. Government. The U.S. and U.K. Governments are authorized to reproduce and distribute reprints for Government purposes notwithstanding any copyright notation hereon. Jorge Lobo was partially supported by the Spanish Ministry of Economy and Competitiveness under Grant Numbers: TIN‐2016‐81032‐P, MDM‐2015‐052, and the U.S. Army Research Office under Agreement Number W911NF1910432.
\bibliographystyle{aaai}
\bibliography{references}
\bigskip
\end{document}